\documentclass[twocolumn,preprintnumbers,amsmath,amssymb,prl]{revtex4}
\usepackage{graphicx}
\usepackage{dcolumn}% Align table columns on decimal point
\usepackage{bm}% bold math
\usepackage{color}

\begin{document}

\title{Mesoscopic simulations of the counterion-induced electroosmotic flow - a
comparative study}

\author{Jens Smiatek$^{1}$}
\altaffiliation{Present address: Institute of Physical Chemistry,
University of M\"unster, Corrensstra\ss e 28/30, D-48149 M\"unster, Germany}
\email{smiatek@physik.uni-bielefeld.de}
\author{Marcello Sega$^2$}
\altaffiliation{Present address: Department of Physics, University of Trento,
and INFN, via Sommarive 14, 38100 Trento, Italy}
\author{Christian Holm$^{2,3}$}
\altaffiliation{Present address: Institut f\"ur Computerphysik,
Universit\"at Stuttgart, Pfaffenwaldring 27, 70569 Stuttgart, Germany}
\author{Ulf D. Schiller$^3$}
\altaffiliation{Present address: Chemical Engineering Department,
University of Florida, Gainesville, FL~32611-6005, USA, {\em on leave
from} Institut f\"ur Festk\"orperforschung, Forschungszentrum J\"ulich,
D-52425 J\"ulich, Germany}
\author{Friederike Schmid$^{1,4}$} 
\affiliation{$^1$ Physics Faculty, Universit{\"a}t Bielefeld, D-33615 Bielefeld, Germany\\
  $^2$ Frankfurt Institute for Advanced Studies, D-60438 Frankfurt, Germany\\
  $^3$ Max-Planck-Institute for Polymer Research, D-55128 Mainz, Germany \\
  $^4$ Institut f\"ur Physik, Universit\"at Mainz, D-55099 Mainz, Germany
}
\date{\today}% It is always \today, today,
             %  but any date may be explicitly specified

% PACS, the Physics and Astronomy
                                   % Classification Scheme.
\begin{abstract}
We present mesoscopic simulations of the counterion-induced
electroosmotic flow in different electrostatic coupling regimes. Two
simulation methods are compared, Dissipative Particle Dynamics (DPD)
and coupled Lattice-Boltzmann/Molecular Dynamics (LB/MD). A general
mapping scheme to match DPD to LB/MD is developed.  For the
weak-coupling regime, analytic expressions for the flow profiles in
the presence of partial-slip as well as no-slip boundary conditions
are derived from the Poisson-Boltzmann and Stokes equations, which are
in good agreement with the numerical results.  The influence of
electrofriction and partial slip on the flow profiles is discussed.
\end{abstract}
\maketitle
\section{Introduction}
\label{sec:introduction}
Microfluidic devices like bio-MEMS (micro-electronical-mechanical-systems) and
bio-NEMS (nano-electronical-mechanical-systems) have attracted broad interest
over the last years due to their huge potential in biotechnology \cite{Harrison,Reyes}. 
The flow profiles in such micro- or nanosized devices are strongly influenced 
by the properties of the boundaries due to the large surface-to-volume ratio 
in these systems.  Surface characteristics like the wetting behavior
and/or slippage have a dramatic effect on the microscopic flow, leading to 
sometimes unexpected behavior \cite{Bocquet}. 

One particularly important mechanism is electroosmotic transport: in contact 
with a liquid, many materials commonly used in nanotechnology ({\em e.g.}, 
polydimethylsiloxane (PDMS)) become charged due to ionizations of surface 
groups \cite{Israelachvili}. As a consequence, surfaces are often covered by
a compensating counterion layer \cite{Hunter}. In an external electric field, the 
ions are driven in one direction, dragging the surrounding solvent with them.
As a result, a flow is induced in the fluid, the electroosmotic flow (EOF). 
This electrokinetic effect has numerous consequences. For example, it alters 
drastically the migration dynamics of mesoscopic objects like polyelectrolytes 
or colloids \cite{Viovy00}. In microchannels, the EOF generated at the channel walls 
results in a total net flow, which is technologically attractive because it
can be controlled and manipulated more easily on the submicrometer scale 
than pressure- or shear-driven flow.

As analytical predictions for flow in such complex systems are often hard to 
derive \cite{Hunter}, numerical simulations are used to investigate it in detail.
The requirements on computer simulations are high: They must include the full 
hydrodynamics of the fluid while guaranteeing, at the same time, optimal
computational efficiency. This has led to the development of a number of 
coarse-grained mesoscopic simulation schemes. Several methods like 
Lattice Gas Automata \cite{Frisch}, the Lattice-Boltzmann Method (LB) 
\cite{McNamara,Benzi,Chen}, Dissipative Particle Dynamics (DPD) 
\cite{Hoogerbrugge,Espanol,Groot}, and Multi-Particle Collision Dynamics 
(MPC) \cite{Malevanets1,Malevanets2} are nowadays used to model liquids at a
coarse-grained level. A completely discretized approach to study EOF {\it via} a
generalized LB method has been proposed by Warren \cite{Warren97a}, and later
extended by Capuani and coworkers \cite{Capuani04a}, relying on a solution of
the electrokinetic equations on the LB lattice.

Compared to atomistic Molecular Dynamics simulations, 
these approaches give access to much longer time- and length scales \cite{Frenkel} 
and are therefore suited to study the long-time behavior of soft matter systems and 
transport phenomena. 
Although the theoretical background of these methods is well understood, their
lattice/off-lattice and thermal/athermal character impedes a straightforward 
mapping between them. 
Comparative studies of specific soft matter problems 
with different simulation methods are therefore scarce.
In this paper we present the results of DPD- and coupled Lattice-Boltzmann/Molecular
Dynamics (LB/MD) simulations for the counterion-induced EOF without salt ions in a slit pore. 
Although this geometry does not allow to test the full coupling between ions and solvent, it has the advantage of possessing an analytical solution, 
which allows a precise test of the accuracy of the methods.
A systematic approach to match DPD and LB/MD simulations is presented.

As mentioned earlier, flow profiles depend heavily on the boundary conditions at 
the surfaces \cite{Bocquet}. We derive an analytic equation for the counterion-induced 
EOF in the presence of no-slip as well as partial-slip boundaries for the 
regime of ``weak electrostatic coupling'', the regime where the Poisson-Boltzmann 
theory is valid, and compare it to our simulation results. Furthermore, we also 
study the influence of the discrete character of wall charges, compared to
perfectly homogeneous walls.

The paper is organized as follows. In the first section we derive the analytical
solution for the flow profile of the counterion-induced EOF in the presence of
arbitrary slip conditions. The two mesoscopic simulation methods are introduced 
and the general mapping scheme is discussed in section III. Section IV gives the
details of the simulations and section V focuses on the numerical results. 
We conclude with a brief summary.

\section{Theoretical Background}
\label{sec:background}
When brought in contact with a liquid, most materials acquire charges either by the 
ionization or dissociation of surface groups or the adsorption of ions from solution. 
In microchannels, the walls are thus typically charged and attract a layer
of compensating counterions. In the following, we consider the simple situation of 
fluid flow in a slit channel with charged walls and dissolved counterions. Additional 
charges ({\em e.g.}, salt ions) are not present. Depending on the relative strength
of Coulomb interactions in the system, one distinguishes between several electrostatic 
coupling regimes, which can be realized by changing the surface charge 
densities, by using counterions of different valency, or by varying the temperature. 
One parameter that discriminates between regimes is the electrostatic coupling 
constant 
\begin{equation}
  \Xi=2\pi Z^3\lambda_B^2\sigma_A,
\end{equation}
which gives the strength of
electrostatic interactions between the counterion and the surface compared
to the thermal energy. The parameters entering $\Xi$ are the Bjerrum length 
$\lambda_B=e^2/4\pi\epsilon_r k_BT$, the unit charge $e$, the counterion valency $Z$, 
the thermal energy $k_BT$, the dielectric constant $\epsilon_r$, and the surface charge 
density $\sigma_A$ \cite{Netz,Moreiraa,Moreirab}. Low values of $\Xi$ correspond to 
the  ``weak coupling limit'', where the counterions at the wall are distributed
in a broad diffusive layer which is well described by the Poisson-Boltzmann 
theory \cite{Gouy,Chapman}. This regime is realized at moderate surface charge
densities and for monovalent ions. High values of $\Xi$ are obtained for high
surface charge densities, low temperatures, or multivalent charges, and lead
to the formation of nearly flat, highly adsorbed and massively correlated counterion 
layers \cite{Boroudjerdi}. If in addition the two counterion layers are well 
separated, {\em i.e.}, the slit width $d$ is much larger than the lateral distance 
between ions $a$ ($d/a > 1$ \cite{Moreiraa}), one enters the ``strong coupling
limit'', which in terms of the electrostatic coupling constant is 
given for $\Xi > (d/\mu)^2$ with the Gouy-Chapman length $\mu=(2\pi Z\lambda_B\sigma_A)^{-1}$.

In the following we focus on the analytical derivation of the EOF profiles in the 
weak coupling regime. The electrostatic potential $\psi(z)$ and the ion density 
distribution $\rho(z)$ can be calculated from the Poisson-Boltzmann equation 
\cite{Gouy, Chapman}, which reads in our case (slit channel, no salt, counterions only)
\begin{equation}
\label{eq:PB}
  \frac{\partial^2}{\partial z^2}\psi(z)=
  -\frac{Ze}{\epsilon_r}\rho(z)=-\frac{Ze}{\epsilon_r}\rho_0 e^{-\frac{Ze\psi(z)}{k_BT}}
\end{equation}
where $Ze$ is the charge of the ions, and $\rho_0$ the reference ion density at $\psi = 0$.
If an additional electric field $E_x$ is applied, the counterions move in the 
direction of the field and drag the solvent particles along. A net flow is induced, 
the counterion-induced EOF. The actual flow profile in the limit of low Reynolds numbers 
can be calculated with the Stokes equation, which reads in absence of pressure gradients
\begin{equation}
  \label{eq:Stokes}
  \eta\frac{\partial^2}{\partial z^2}v_x(z) =Ze \rho(z) \: E_x
\end{equation} 
with the dynamic viscosity $\eta$. We impose the most general hydrodynamic boundary 
condition, the partial slip condition
\begin{equation}
  \label{eq:slip}
  \partial_z v_x(z)|_{z=\pm z_B}=\mp \frac{1}{\delta_B}v_x(\pm z_B) 
\end{equation}
at the hydrodynamic wall boundaries $\pm z_B$ with the slip length $\delta_B$. 
Comparing Eq.~(\ref{eq:Stokes}) with Eq.~(\ref{eq:PB}), we obtain by
straightforward integration of Eq.~(\ref{eq:Stokes}) with (\ref{eq:slip})
the general relation
\begin{equation}
  \label{eq:vx}
   v_x(z) = \frac{\epsilon_r E_x}{\eta} ( \psi(z_B)-\psi(z) - \delta_B \psi'(z_B)).
\end{equation}
Specifically, the solution of Eq.~(\ref{eq:PB}) in the slit geometry is given 
by~\cite{Israelachvili}
\begin{equation}
  \label{eq:psi}
  \psi(z)=\frac{k_BT}{Ze}\log(\cos^2(\kappa z))
\end{equation}
with the screening constant $\kappa^2=(Ze)^2\rho_0/2\epsilon_r k_BT$,
where $\rho_0$ is the counterion density in the middle of the channel. 
The counterion density distribution is then given by
\begin{equation}
  \label{eq:rho}
  \rho(z)=\frac{\rho_0}{\cos^2(\kappa z)},
\end{equation}
and the electric field is
\begin{equation}
  \label{eq:E_z}
  E(z)=-\frac{\partial}{\partial
    z}\psi(z)=-\frac{2\kappa k_BT}{Ze}\tan(\kappa z).
\end{equation}
For the flow profile, we finally get
\begin{eqnarray}
  \label{eq:ana_EOF}
  v_x(z) &=&
  \frac{e }{4 \pi \lambda_B Z \eta}E_x(\log(\cos^2(\kappa z_B))\nonumber\\
   & & -\log(\cos^2(\kappa z)) + 2 \kappa \delta_B\tan(\kappa z_B)),
\end{eqnarray}
where we have expressed $\epsilon_r$ in terms of the Bjerrum length $\lambda_B$.

If the surface charges are high, the ions have high valency, or the temperature is low, 
charge correlations and charge fluctuations become important and mean-field approaches 
like the Poisson-Boltzmann approach fail. In this case, asymptotic analytical expressions 
for the ion distribution in the channel can be obtained with field-theoretic 
methods\cite{Netz,Moreiraa,Moreirab}. If the plate distance $d$ is much 
larger than the Gouy-Chapman length, 
the counterion density distribution between two planar 
highly charged surfaces is given by \cite{Moreiraa} 
\begin{equation}
  \label{eq:SC_density}
 \rho(z)=\frac{2\pi\lambda_B\sigma_A^2}{(1-e^{-d/\mu})}\left(e^{-(z+d/2)/\mu}+e^{-(d/2-z)/\mu}\right).
\end{equation}
This result characterizes the ``strong coupling limit'', corresponding to two independent 
highly self-correlated counterion layers.

\section{Simulation methods}
\label{sec:SM}
\subsection{Dissipative Particle Dynamics}
\label{sec:DPD}

Dissipative Particle Dynamics (DPD) is a coarse-grained particle-based
simulation method for fluid dynamics. It is Galilean-invariant,
generates a well-defined canonical ensemble at equilibrium
and guarantees the conservation 
of momentum. The system consists of a set of particles 
with continuous positions ${\bf{r}}_i$ and velocities ${\bf{v}}_i$ whose time 
evolution is described by Newton's equations of motion. The forces on one 
particle are given by
\begin{eqnarray}
  {\bf{F}}_{i}^{DPD}=  
  \sum_{i\not={j}} \Big({\bf F}_{ij}^C+ {\bf{F}}_{ij}^{D}+{\bf{F}}_{ij}^{R} \Big)
\end{eqnarray}
with the conservative force ${\bf F}_{ij}^C$, a two-particle dissipative interaction 
\begin{eqnarray}
  {\bf{F}}_{ij}^{D} =
  -\gamma\omega(r_{ij})(\hat{r}_{ij}\cdot{\bf{v}}_{ij})\hat{r}_{ij}
\end{eqnarray}
with $\hat{r}_{ij} = {\bf r}_{ij}/r_{ij}$, ${\bf r}_{ij} = {\bf r}_j - {\bf r}_i$,
and a corresponding stochastic force ${\bf{F}}_{ij}^{R}$  
\begin{eqnarray}
  {\bf{F}}_{ij}^{R} = \sqrt{2 \gamma k_B T \omega(r_{ij})} \hat{\zeta}_{ij}\hat{r}_{ij}.
\end{eqnarray} 
The weight function $\omega(r)$ can be chosen at will -- here we use the most common 
form $\omega(r) = 1-r/r_c$ for $r < r_c$ ($\omega(r) = 0$ otherwise), with the DPD cut-off 
radius $r_c$. The random numbers $\hat{\zeta}_{ij}$ are symmetric and otherwise uncorrelated 
($\hat{\zeta}_{ij}=\hat{\zeta}_{ji}$) with mean zero and variance one. 
The symmetry property $F_{ij}^D = - F_{ji}^D$, and $F_{ij}^R = - F_{ji}^R$ 
ensures that the total momentum is conserved in the absence of spatially varying 
external (conservative) potentials.\\
The hydrodynamic boundary condition at the walls (\ref{eq:slip}) is
realized with a recently developed method \cite{Smiatek} that allows
to implement arbitrary partial-slip boundary conditions:
We introduce an additional coordinate-dependent viscous force 
that mimicks the wall/fluid friction
\begin{equation}
\label{eq:langevin}
  {\bf{F}}_i^{L}={\bf{F}}_i^{D}+{\bf{F}}_i^{R}
\end{equation}
with a dissipative contribution
\begin{equation}
  {\bf{F}}_i^{D}=-\gamma_L\: \omega_L(z) \: \: ({\bf{v}}_i-{\bf{v}}_{wall})
\end{equation}
coupling to the relative velocity $({\bf{v}}_i-{\bf{v}}_{wall})$ of the particle
with respect to the wall, and a stochastic force
\begin{equation}
  F_{i,\alpha}^R= \sqrt{2 \gamma_L \: k_B T \: \omega_L(z)}\;\chi_{i,\alpha}
\end{equation}
which satisfies the fluctuation-dissipation relation and thus ensures the
Boltzmann distribution to be the local equilibrium distribution.
Here $\alpha$ is $x,y,z$ and $\chi_{i,\alpha}$ is a Gaussian distributed
random variable with mean zero and variance one: $\langle \chi_{i,\alpha} \rangle = 0$,
$\langle \chi_{i,\alpha} \chi_{j,\beta} \rangle = \delta_{ij} \delta_{\alpha \beta}$.
The viscous coupling between fluid and wall is realized by the locally varying 
viscosity $\gamma_L \omega_L(z)$ with $\omega_L(z) = 1 - z/z_c$ up to a cut-off 
distance $z_c$. Beyond $z \le z_c$, the fluid satisfies the Stokes equation
with an effective boundary condition of the form (\ref{eq:slip}) \cite{Smiatek}.
The prefactor $\gamma_L$ can be used to tune the strength of the friction
force and hence the value of the slip length $\delta_B$. We note that 
$\delta_B$ is an effective parameter and not related to the actual slip
at the physical walls, which is usually nonzero even at $\delta_B=0$. 
Within this approach it is possible to tune the hydrodynamic boundary condition
systematically from full-slip to no-slip, and to derive an analytic expression 
for the slip length as a function of the model parameters \cite{Smiatek}.
Theorists tend to favor no-slip boundary conditions. In microchannels,
however, partial-slip boundary conditions are also observed \cite{Pit,Tretheway,Neto}.  
Here we will show results for both no-slip and partial-slip walls.

\subsection{Lattice-Boltzmann Approach}
\label{sec:LBM}
In contrast to DPD, the Lattice-Boltzmann (LB) method can be seen as a
discrete formulation of the Boltzmann equation on a lattice, which, by
means of a Chapman-Enskog expansion leads to the Navier-Stokes
equation in the incompressible limit \cite{Succi,Duenweg}.  The basic
evolution equation for the mass density $n_i({\bf{r}},t)$ assigned to
a discrete velocity ${\bf{c}}_i = \hat{\bf{c}}_i a/\tau$ on a lattice
node ${\bf{r}}$ at time $t$ is given by
\begin{equation}
  n_i({\bf{r}}+{\bf{c}}_i{\tau},t+\tau)=
  n_i({\bf{r}},t)+\sum_j L_{ij}\left(n_j({\bf{r}},t)-n_j^{eq}(\rho,{\bf{u}})\right) ,
\end{equation}
where $\tau$ is the time step, and $\hat{\bf{c}}_i$ is a set of
vectors connecting the grid point ${\bf{r}}$ to its nearest and
next-nearest neighbors on a simple cubic lattice with lattice spacing
$a$. The zero-velocity vector is also included (D3Q19 model).
The collision matrix $L$ relaxes the $n_i$ towards the local
pseudo-equilibrium $n_i^{eq}(\rho,{\bf{u}})$, which depends on the
local mass density $\rho({\bf{r}},t) = \sum_i n_i({\bf{r}},t)$ and the
local fluid velocity ${\bf{u}}({\bf{r}},t) = (1/\rho) \sum_i
n_i({\bf{r}},t) {\bf{c}}_i$.  The functional form of the
pseudo-equilibrium is taken as the second-order approximation
\begin{equation}
n_i^{eq}(\rho,{\bf{u}}) = \rho a^{c_i} \left( 1+ \frac{{\bf{u}} \cdot
{\bf{c}}_i}{c_s^2} + \frac{({\bf{u}} \cdot {\bf{c}}_i)^2}{2c_s^4} -
\frac{u^2}{2c_s^2} \right) ,
\end{equation}
which maximizes the entropy of the underlying generalized lattice gas
model \cite{Duenweg}. $c_s$ is the speed of sound and the parameters
$a^{c_i}$ are weight factors that depend on the magnitude of the
velocity vectors $\hat{\bf{c}}_i$ but not their direction. For the
D3Q19 model the respective values are $a^0=1/3$, $a^1=1/18$ and
$a^{\sqrt{2}}=1/36$, and the speed of sound is $c_s=\sqrt{1/3} \
a/\tau$.
In order to reproduce Brownian motion in a suspension, thermal
fluctuations have to be introduced. The fluctuating lattice Boltzmann
equation is given by
\begin{eqnarray}
  n_i({\bf{r}}+{\bf{c}}_i{\tau},t+\tau) &=&
  n_i({\bf{r}},t)+\nonumber\\
& & \sum_j
  L_{ij}\left(n_j({\bf{r}},t)-n_j^{eq}(\rho,{\bf{u}})\right)\nonumber\\
& & +n_i^{\prime}({\bf{r}},t).
\end{eqnarray}
For details regarding the implementation of the stochastic term
$n_i^{\prime}({\bf{r}},t)$ we refer to Ref. \onlinecite{Duenweg}.
A delicate task in lattice-based simulations is the correct coupling
between the discrete nodes of the LB-solvent and the continuous
positions of the (ion) particles. This can be achieved with a
Stokes-like friction force \cite{Ahlrichs1}
\begin{equation}
  \label{eq:coupling}
  {\bf{F}}_{ml}=-\zeta[{\bf{V}}-{\bf{u}}({\bf{R}},t)]+{\bf{f}} ,
\end{equation}
which is exerted on a solute particle moving through the surrounding
viscous fluid.  The random force ${\bf{f}}$ is required to guarantee
thermal equilibrium of the coupled system, and its amplitude can be
obtained from the fluctuation dissipation relation
$<f_{\alpha}(t)f_{\beta}(t')>=2\zeta
k_BT\delta_{\alpha\beta}\delta(t-t')$. 
Momentum is exchanged between the particles and the LB fluid according
to total momentum conservation in the system.
The LB simulations were carried out for no-slip boundary conditions
only, which were realized by applying bounce-back boundary conditions \cite{Revenga}.

\subsection{Parameter Mapping}
\label{sec:pm}
Our scheme for mapping the parameters of the different simulation methods onto each
other is based on the requirement that the hydrodynamic flow phenomena should be
the same. Therefore, the values of the parameters entering the Stokes equation
(\ref{eq:Stokes}) should be identical, {\em i.e.}, the solvent density $\rho$ and
the dynamic viscosity $\eta$. Matching the solvent density $\rho$ is trivial, because
it is an input parameter both in the DPD and the LB method. Matching the dynamic
viscosity $\eta$ is more difficult. This parameter is an input parameter in the LB
method, but an {\em a priori} unknown intrinsic fluid property in DPD models. A 
mean-field analysis of the DPD method \cite{Marsh} shows that it mainly depends on the friction 
coefficient $\gamma$, the fluid density $\rho$, the temperature $T$, and the cut-off 
range of DPD interactions $r_c$. For given $\rho$ and $T$, it can hence be adjusted
by varying $\gamma$ and/or $r_c$. The dynamic viscosity can be determined either by measuring
the autocorrelation function of the pressure tensor in free solution, which is related 
to the viscosity {\em via} a Green-Kubo equation \cite{Allen}, or by simulating
Plane Poiseuille flow in a confined microgeometry \cite{Smiatek}, respectively
in free
periodic boxes \cite{Backer}. The resulting flow profile is then fitted to
a parabola $v_x(z) = \alpha (z_p^2-z^2)$ with the magnitude
\begin{equation}
\label{eq:alpha}
  \eta=\frac{\rho F_x}{2\alpha},
\end{equation}
where $F_x$ is the body force on the solvent which was applied to generate
the Poiseuille flow.
Having fixed $\rho$ and $\eta$, a third parameter remains which is related
to the long-time dynamical behavior of single particles, {\em i.e.}, to the
self-diffusion. This is another intrinsic fluid property in DPD fluids.
The LB method does not operate with particles originally, but in the hybrid
LB/MD-method (see section III.B), particles can be introduced which 
are coupled to the LB-nodes with a coupling constant $\zeta$ 
(cf. Eq.~(\ref{eq:coupling})). Therefore the self-diffusion coefficient $D$ 
can be tuned by varying the coupling constant $\zeta$ until it is in
agreement with the self-diffusion coefficient of a solute particle in the DPD simulations.
This can be checked by comparing the velocity autocorrelation function of a
particle in both systems which is connected to the self-diffusion
coefficient $D$ by a Green-Kubo expression, or by simple comparison of the
mean-square displacement of a tracer particle \cite{Allen}.

\section{Simulation details}
\label{sec:Details}

All simulations have been carried out with extensions of the freely available
software package {\sf{ESPResSo}} \cite{Espresso1,Espresso2,Espresso3}.  
We use a cubic simulation box ($12\sigma\times 12\sigma\times 12\sigma$) 
which is periodic in the $x$- and $y-$direction and confined by impermeable 
walls in the $z$ direction. Electrostatics for homogeneously charged walls
are calculated by P3M \cite{Hockney} 
and the ELC-algorithm \cite{Arnold} for $2D+h$ slab geometries with
an ELC gap size of $2 \sigma$. See Ref.\onlinecite{Deserno} for a definition of
the parameters. In addition, the MMM2D algorithm
\cite{Arnold02b,Arnold02c} was used to study homogeneously charged walls in the
same geometry. Ions have a mass $m$ and carry a 
charge $q$. They interact {\em via} a Weeks-Chandler-Andersen(WCA)-potential \cite{WCA} 
$V_{ij}=4\epsilon((\sigma/r)^{12}-(\sigma/r)^{6})+\epsilon$ 
with energy parameter $\epsilon$ and cut-off distance 
$r_{WCA}=2^{1/6}\sigma$, and a Coulomb potential with Bjerrum 
length $\lambda_B=1.0\sigma$. The parameters $\epsilon$, $\sigma$, 
and $m$ shall be used as the natural units of our system hereafter.
To study the counterion-induced electroosmotic flow, the walls were 
charged by placing discrete charges $q$ randomly all over them.
The charge configuration in each wall was identical for both methods 
with a surface ion density $\sigma_{s}$ (the surface {\em charge} density
is then given by $\sigma_A = q\sigma_s$). The number
of counterions in the fluid matches the number of charges in the
wall, such that the system is overall neutral in the case of inhomogeneously
charged walls. The case of homogeneously charged walls can be realized 
simply by  having both walls uncharged. This is due to the fact that the 
electric fields generated by the two plates cancel each other exactly within 
the slit, and the charge density profile is generated only by the 
internal Coulomb interaction between counterions. 

Different coupling regimes were obtained by varying the ion charge $q$, the surface ion density
$\sigma_s$ and the corresponding counterion number from $12$ to $60$. 
\begin{table}
 \caption{\label{tab1} Charge $q$, surface ion density $\sigma_s$,
      Gouy-Chapman length $\mu$ and 
      coupling constant $\Xi$ for different electrostatic coupling regimes.}
    \begin{tabular}{|l|c|c|c|c|}
    \hline
    Regime & $q[\sqrt{k_BT\lambda_B}]$ & $\sigma_s[1/\sigma^2]$ & $\mu
    [\sigma]$ & $\Xi$ \\
    \hline
    Weak coupling & 1 & 0.2083 & 0.764 & 1.31 \\
    \hline
    Intermediate coupling & 2 & 0.042 & 0.955 & 4.19 \\
    \hline
  \end{tabular}
\end{table}
The parameters are shown in Table \ref{tab1} together with the resulting
coupling constant $\Xi$. The density of solvent particles was chosen 
$\rho_l=3.75\sigma^{-3}$, and the temperature $k_B T = 1 \epsilon$ in all simulations.

\subsection{Dissipative particle dynamics simulations}

In the DPD simulations, the walls are located at $z_{wall}=(0,10)\sigma$. 
They interact with fluid particles and ions 
{\em via} a WCA-potential \cite{WCA} with the same parameters as above
($\epsilon_{wall}=\epsilon, \sigma_{wall}=\sigma$). The solvent particles have
the same mass as the ions ($m$), but do not interact with other particles
except the walls. The DPD friction coefficient was chosen
$\gamma=5.0\sigma^{-1}(m\epsilon)^{1/2}$ and the cutoff range of the DPD
interactions $r_c=1.0\sigma$. Tunable slip boundary conditions \cite{Smiatek}
were used with friction coefficients $\gamma_L=0.96, 1.4049, 3.1 \sigma^{-1}(m\epsilon)^{1/2}$
and $\gamma_L=6.1\sigma^{-1}(m\epsilon)^{1/2}$ 
with the friction range $z_c=2.0\sigma$ starting at $z_{wall}=(0,10)\sigma$.
The DPD timestep is $\delta t=0.01\sigma(m/\epsilon)^{1/2}$. 

\subsection{Lattice Boltzmann simulations}

The LB simulations were carried out using the D3Q19 model with $24^3$
solvent nodes. The walls for
the ions are placed as in the DPD simulations at
$z_{wall}=(0,10)\sigma$. Since the zero-velocity surface in a LB simulation is located, 
at first order in viscosity, halfway in between two rows of nodes, a shift has been added to the
WCA wall potential in order to match the position of the zero-potential and zero-velocity planes.
The grid spacing of the LB fluid is $a=0.5\sigma$. The coupling constant of the fluid with the
ions is $\zeta=1.98\sigma^{-1}(m\epsilon)^{1/2}$ which was derived by the
mapping scheme proposed in Section 5. The integration  timestep for the ions 
as well as for the fluid was $\delta t=\tau = 0.01\sigma(m/\epsilon)^{1/2}$, 
bounce-back boundary conditions were applied on the fluid at the wall
positions to create no-slip boundary conditions.

\section{Results}
\label{sec:results}
\subsection{Computational cost}
An important criterion when choosing a mesoscopic simulation method is its computational cost 
and efficiency. We have measured the time to calculate a single time step in both methods on an 
Athlon${\tiny{^{\copyright}}}$ MP2200+ CPU. 
\begin{table}
  \caption{\label{tab2} Time needed for computing a single integration step
      in the DPD method with tunable-slip boundary conditions (DPD+TSC) and
      electrostatics (DPD+TSC+CS) in comparison to the LB method with bounce-back
      boundary conditions (LB+BBC) and electrostatics (LB+BBC+CS).}
  \begin{tabular}{|l|c|c|c|c|}
    \hline
    Methods & DPD & LB & DPD+CS & LB+CS\\
    \hline
    Time/step [s] & 0.09 & 0.01 & 0.22 & 0.14 \\
    \hline
  \end{tabular}
\end{table}
The values are presented in Table \ref{tab2}.
The first two columns show the time spent on an uncharged system with $4320$ 
solvent particles (DPD) with tunable-slip boundary conditions (TSC) or $1728$
solvent nodes (LB) in the above mentioned 
microgeometry with bounce-back boundary conditions (BBC). The LB-method is nine times faster 
than the DPD-method. We should note, however, that these values strongly depend on 
the density of solvent particles, or the number of solvent nodes, respectively. 
The last two columns show the corresponding values for a charged system (CS)
with the P3M in combination with the ELC algorithm for discretely charged walls. 
It is evident that most of the time is spent on the calculation of the 
electrostatic interactions for 60 ions and 60 counterions ($\sigma_s=0.208\sigma^{-2}$). 
If electrostatic interactions 
are considered, both methods are comparable with respect to their computational 
cost and efficiency, although it has to be noticed that the efficiency of P3M
strongly depends on the chosen parameter values (Ewald parameter
$\alpha=2.1875$, mesh size $32^3$). Depending on the choice of parameter and of 
system size the computational load of computing electrostatic calculation can be
even lower than that of hydrodynamics.

\subsection{Fluid properties}
We first consider the dynamic properties of the bulk fluid.  To measure the dynamic 
viscosity of the DPD fluid, which is needed for the parameter mapping, we
fitted a plane Poiseuille flow as in Ref.~\onlinecite{Smiatek}. This procedure yields 
the value $\eta=(1.334\pm0.003)\sigma^{-2}(m\epsilon)^{1/2}$ both for uncharged
and charged fluids.
\begin{figure}[t]
  \includegraphics[width=8.5cm]{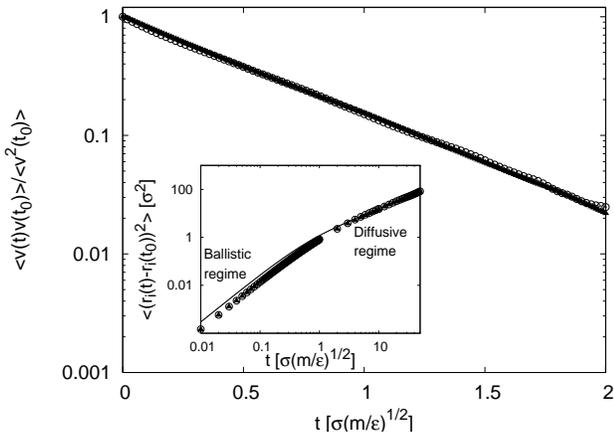}
  \caption{Normalized velocity autocorrelation function for a DPD-fluid
    particle (open circles) with number density $\rho=3.75{\sigma^{-3}}$, DPD friction 
    coefficient $\gamma=5.0\sigma^{-1}(m\epsilon)^{1/2}$ and for a LB/MD-fluid particle 
    (filled triangles) with the same density and coupling constant 
    $\zeta=1.98\sigma^{-1}(m\epsilon)^{1/2}$ .
    The characteristic decay time for the DPD-method is
    $\tau_{DPD}=(0.5162\pm0.0008)\sigma(m/\epsilon)^{1/2}$ and for the
    LB-method $\tau_{LB}=(0.5218\pm0.0006)\sigma(m/\epsilon)^{1/2}$.
    Inset: Mean square displacement for a fluid DPD particle (open circles) and
    for a coupled LB/MD particle (filled triangles) compared to Eq.~\ref{eq:MSD}
    (solid line). 
  }
  \label{fig:vv_msd}
\end{figure}
Next we calculated the effective diffusion coefficient $D$ for a single tracer particle 
in the DPD-method and matched it to the tracer diffusion in the LB-model.
The self-diffusion coefficient $D$ can be obtained by exploiting a Green-Kubo
expression \cite{Allen}, which relates $D$ to the velocity autocorrelation function 
(see Fig.~\ref{fig:vv_msd})
\begin{equation}
  \label{eq:vv_msd}
  D=\frac{1}{3}\int_{t_0}^{\infty}dt<{\bf{v}}(t){\bf{v}}(t_0)>.
\end{equation}
Fig.~\ref{fig:vv_msd} shows that the velocity autocorrelation function decays
exponentially over at least two orders of magnitude. 
Approximating the integrand in Eq.~(\ref{eq:vv_msd}) by this exponential law
we obtain $D_{DPD}=(0.2581\pm0.0004)\sigma(m/\epsilon)^{-1/2}$ for the particle in
the DPD fluid (at late times, one expects to see an algebraic long-time tail in the
velocity autocorrelation function, but its contribution to $D$ is small and well within 
the error). A coinciding value of the self-diffusion coefficient in the LB/MD-method can 
be obtained by setting the coupling constant to $\zeta=1.98\sigma^{-1}(m/\epsilon)^{1/2}$. 
The two corresponding velocity autocorrelation function then closely match each other
(Fig.~\ref{fig:vv_msd}) and the resulting self-diffusion coefficient for LB/MD-fluid
particles is given by $D_{LB}=(0.2609\pm0.0003)\sigma(m/\epsilon)^{-1/2}$.

Alternatively, the self-diffusion coefficient can be determined directly from
the mean-square displacement of a single solvent particle at late times
\begin{equation}
  D=\lim_{t\to\infty} \frac{<({\bf{r}}_i(t)-{\bf{r}}_i(t_0))^2>}{6t}.
\end{equation}
The results for the mean-square displacement are shown in the inset of Fig.~\ref{fig:vv_msd}. 
In agreement with standard theories \cite{Paul,vanKampen}, ballistic behavior 
($\sim t^2)$ is observed at short times ($t \leq 0.75 \sigma(m/\epsilon)^{1/2}$), 
but diffusive behavior dominates after a characteristic time which is 
roughly $t\geq 10\sigma(m/\epsilon)^{1/2}$ for our model parameters.
A linear regression of the diffusive regime yields 
$D_{DPD}=(0.2698\pm0.0002)\sigma(m/\epsilon)^{-1/2}$ and
$D_{LB}=(0.2617\pm0.0005)\sigma(m/\epsilon)^{-1/2}$, which is in 
rough agreement with the Green-Kubo results.
It is remarkable that the numerical results obtained with both methods lie 
almost on top of each other. Assuming that the velocity of a single particle
propagates according to an Ornstein-Uhlenbeck process, the full-time mean-square 
displacement can be calculated analytically as a function of the effective friction 
coefficient $\zeta_{e}=k_BT/D$ and is given by \cite{Paul}
\begin{eqnarray}
\label{eq:MSD}
<({\bf{r}}_i(t)-{\bf{r}}_i(t_0))^2> &=&
6\,\frac{k_BT}{m\zeta_{e}}t+\frac{<{\bf{v}}^2(t_0)>}{\zeta_{e}^2}(1-e^{-\zeta_{e}t})^2\nonumber\\
& & -\frac{k_BT}{m\zeta_{e}^2}(3-4e^{-\zeta_{e}t}+e^{-2\zeta_{e}t}).
\end{eqnarray}
This prediction is shown as the straight line in the inset of Fig.~\ref{fig:vv_msd}.
It is in reasonable agreement with the numerical results.

Finally, we consider uncharged DPD fluids in slit geometry and establish the values of
the slip length $\delta_B$ and the hydrodynamic boundary positions $z_B$ for different 
values of the surface friction parameter $\gamma_L$. By a combination of plane
Poiseuille flow and plane Couette flow simulations it is possible to determine 
$\delta_B$ and $z_B$ independently \cite{Smiatek}.
\begin{table}
  \caption{\label{tab3} 
    Slip lengths for different layer friction coefficients
    $\gamma_L$, determined for uncharged fluid flow. The numerical results are 
    compared to the theoretical prediction of Ref.25.
  }
  \begin{tabular}{|l|c|c|c|c|}
    \hline
    $\gamma_L$ & $0.96$ & $1.4049$ & $3.1$ & $6.1$\\
    \hline
    $\delta_B^m$ & $1.399\pm 0.385$ & $0.782 \pm 0.246$ &
    $0.248 \pm 0.231$ & $0.000 \pm 0.197$\\
    \hline
    $\delta_B^t$ & $1.249$ & $0.780$ & $0.226$ & $0.000$\\
    \hline
  \end{tabular}
\end{table}
The resulting hydrodynamic boundary positions were found to be located at
$|z_B| = (3.866 \pm 0.265)\sigma$, and the corresponding results of the slip
length compared to the theoretical prediction of the analytical expression 
derived in \cite{Smiatek} are shown in Table \ref{tab3}.

Having matched the model parameters in the DPD and the LB/MD method, we can now
proceed to simulate the counterion-induced electroosmotic flow and to compare the 
results.

\subsection{EOF: Weak-coupling regime}
First we consider the counterion distribution in the channel with inhomogeneously 
charged walls in the weak-coupling regime ($\Xi=1.31)$.
\begin{figure}[ht]
    \includegraphics[width=8.5cm]{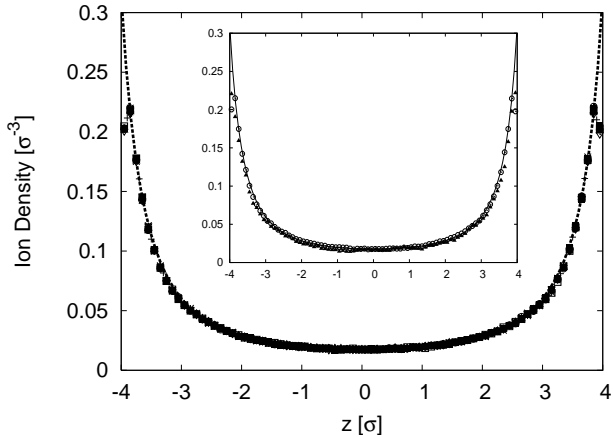}
    \caption{Counterion density distribution for the DPD-fluid in the
      weak-coupling limit ($\Xi=1.31$) between two charged
      walls for external electric field strengths between
      $E_x=0.1-1.0k_BT/e\sigma$. The external perpendicular electric field does
      not perturb the ion density distribution.
      The dashed line presents the theoretical prediction of the
      Poisson-Boltzmann-Theory for an ion density in the middle of the
      channel of
      $\rho_0=(0.0176\pm0.0001)\sigma^{-3}$.
      {\bf Inset:} Comparison of counterion distribution for the DPD-method (circles)
      and for the LB-method (triangles) with a field strength $E_x=1.0k_B T/e\sigma$. The
      straight line is again the theoretical prediction of the
      Poisson-Boltzmann-Theory.
    }
    \label{fig:ion}
\end{figure}
DPD simulations were carried out for ten different external electric field 
strengths between $E_x=0.1-1.0k_BT/e\sigma$. Fig.~\ref{fig:ion} shows that 
the counterion distribution between the walls is not perturbed by the applied 
fields nor the discreteness of charges in the walls in the accessible $z$-range.
For all fields, it is in very good agreement with the theoretical prediction 
of the Poisson-Boltzmann equation (\ref{eq:rho}) with the only fit 
parameter $\rho_0=(0.0176\pm0.0001)\sigma^{-3}$. The inset of
Fig.~\ref{fig:ion} compares the counterion distribution in the DPD-fluid
with the corresponding ion distribution in the LB/MD-fluid for a field strength
$E_x=1.0k_BT/e\sigma$. The simulation results are nearly identical and follow again
closely the prediction of the electrostatic theory.

\begin{figure}[ht]
  \begin{center}
    \includegraphics[width=8.5cm]{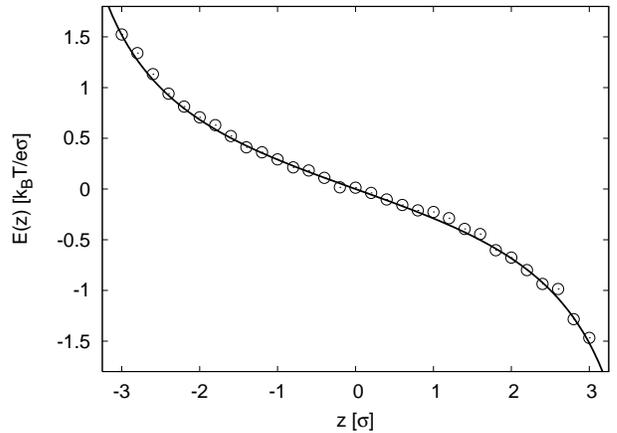}
    \caption{Electric field $E(z)$ in the solvent measured by a test charge
      method in the weak-coupling limit ($\Xi=1.31$). 
      The dashed line shows the theoretical prediction of Eq.~(\ref{eq:E_z}).
    }
    \label{fig:field}
  \end{center}
\end{figure}
In addition, we also measured the electric field $E(z)$ by a test charge method.
The results are shown in Fig.~\ref{fig:field}, they also agree very well with
the prediction of the Poisson-Boltzmann theory, Eq.~(\ref{eq:E_z}).\\
\begin{figure}
    \includegraphics[width=8.5cm]{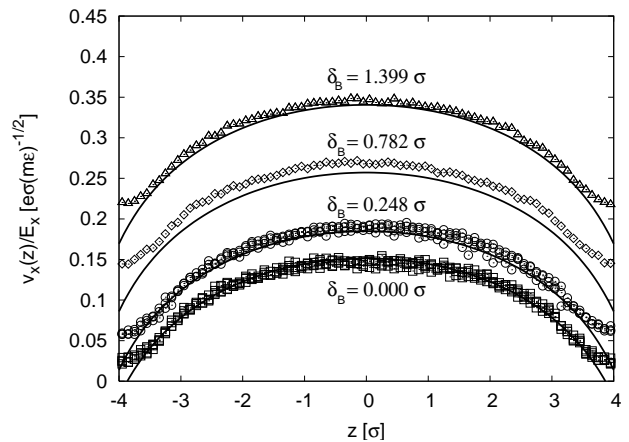}
    \caption{Flow profiles for the DPD- method with various
      field strengths $E_x=0.8-1.0k_B T/e\sigma$ for
      varying slip lengths in the weak coupling regime
      ($\Xi=1.31$). The hydrodynamic boundary positions for the DPD-method
      are at $|z_B|=(3.866\pm 0.265)\sigma$. The straight lines represent the
      theoretical prediction of 
      Eqn.(\ref{eq:ana_EOF}). 
    }
    \label{fig:EOF}
\end{figure}
\begin{figure}
    \includegraphics[width=8.5cm]{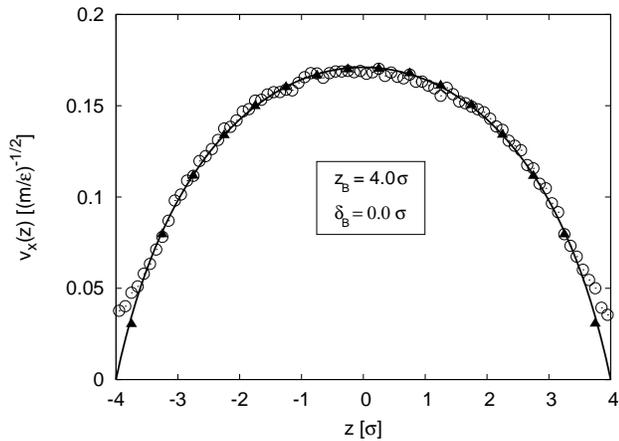}
    \caption{Flow profile for the DPD-(circles) and the LB/MD-method (triangles) with
      no-slip boundary conditions. The straight line is the theoretical
      prediction of Eqn.~(\ref{eq:ana_EOF}) with $|z_B|=4.0\sigma$ and
      $\delta_B=0.0\sigma$.
    }
    \label{fig:comp_EOF}
\end{figure}
In addition, we also measured the electric field $E(z)$ by a test charge method.
The results are shown in Fig.~\ref{fig:field}, they also agree very well with
the prediction of the Poisson-Boltzmann theory, Eq.~(\ref{eq:E_z}).

The influence of the partial-slip boundary conditions is presented in
Fig.~\ref{fig:EOF} for varying field strength $E_x$. In agreement with the
theoretical prediction, Eq.~(\ref{eq:ana_EOF}), we find that $v_x \propto E_x$
for all slip lengths.  Varying the slip length has a quite dramatic effect on 
the magnitude of the flow profiles. 
All numerical results are in reasonable agreement with Eq.~(\ref{eq:ana_EOF}), 
especially if one bears in mind that 
the theoretical curves have an uncertainty due to the numerical error of $\delta_B$. 

The comparison of the DPD- and the LB/MD flow profiles is finally presented 
in Fig.~\ref{fig:comp_EOF} for no-slip boundary conditions. Here we have shifted the 
effective channel width in the DPD method from $8 \sigma$ to $8.28 \sigma$, 
such that the hydrodynamic boundaries were now located at $|z_B|=4.030 \pm 0.357\sigma$ 
like in the LB/MD-fluid.  Fig.~\ref{fig:comp_EOF} shows that the flow profiles 
obtained with both methods are then identical and agree well with the theoretical
prediction, Eq.~(\ref{eq:ana_EOF}), evaluated with $\delta_B=0$ and $|z_B|=4\sigma$.

\subsection{EOF: Homogeneously and Inhomogeneously charged walls}
The results shown up to now have been obtained in system where discrete charges were
placed randomly in the walls.
\begin{figure}
    \includegraphics[width=8.5cm]{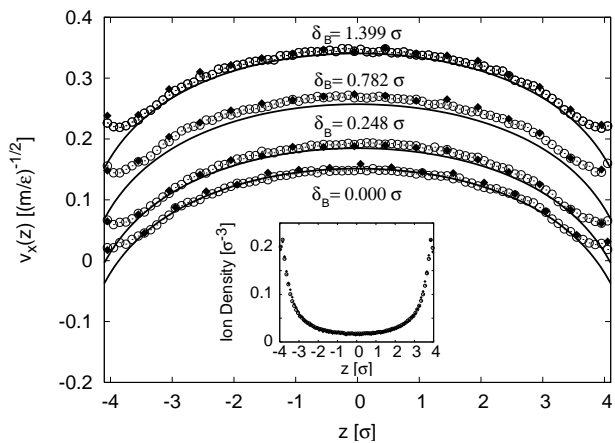}
    \caption{\label{fig:EOF_homo_inhomo}Flow profile for various slip lengths with homogeneously
      (filled diamonds) and inhomogeneously
      charged walls (circles) for $E_x=1.0k_B T/e\sigma$ 
      in comparison to the theoretical prediction of Eqn.~(\ref{eq:slip}).
      {\bf Inset:} Counterion distribution with $\sigma_A=0.208 e\sigma^{-2}$ between homogeneously 
      respectively inhomogeneously charged walls. Only slight deviations can
      be observed.
    }
\end{figure}
 For comparison, we have also studied systems with 
homogeneously charged walls with DPD simulations. Surface charge inhomogeneities 
may lead to electrofriction and slow down the fluid at the wall \cite{Netz03}.
Since the electrostatic potential between two equally and homogeneously charged walls 
is constant, the charges on the wall can be omitted and it suffices to study a fluid
with ions confined between uncharged walls. The simulations were carried out using the 
MMM2D-algorithm \cite{Arnold02b, Arnold02c} and the parameters given in
Table \ref{tab3} for the weak-coupling (Poisson-Boltzmann) regime. 
The ion density $\rho$ was $0.0525\sigma^{-3}$ in all simulations.
Fig.~\ref{fig:EOF_homo_inhomo} presents the numerical results for various slip lengths 
for homogeneously and inhomogeneously charged walls, compared to the theory of 
Eq.~(\ref{eq:ana_EOF}) for an external field of $E_x=1.0 k_B T /e\sigma$. All flow profiles 
for homogeneously respectively inhomogeneously charged walls are identical. 
These results are consistent with theories \cite{Netz03,Joly04} -- the
electrostatic interaction is moderate and drastic deviations have only been reported for
strongly interacting systems \cite{Netz03}. In the weak-coupling regime,
the influence of electrofriction on the flow profiles both for no-slip and
partial-slip boundaries is thus negligible.

\subsection{EOF: Intermediate-coupling regime}
Finally, we consider a situation where no analytical theory is available, and
study the intermediate coupling regime with $\Xi=4.19$.
\begin{figure}[ht]
    \includegraphics[width=8.5cm]{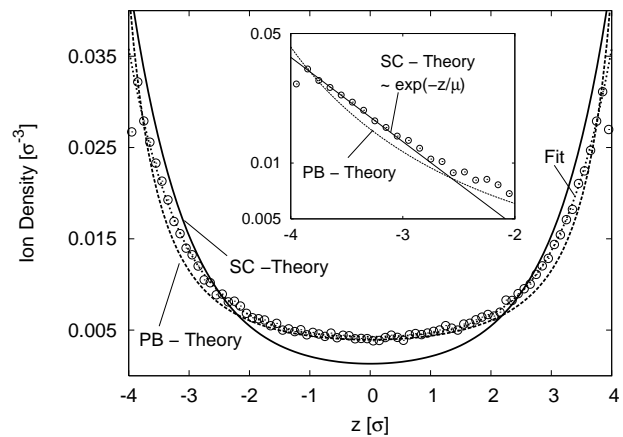}
    \caption{Counterion density distribution with
      $\rho=0.0104\sigma^{-3}$ in the intermediate coupling regime
      with $\Xi=4.19$ for a surface ion density
      $\sigma_s=0.042 \sigma^{-2}$. The straight line shows the theoretical
      prediction of the strong-coupling limit (Eq.(\ref{eq:SC_density})). 
      Neither the strong coupling theory (straight line) nor the
      Poisson-Boltzmann theory (dashed line at the bottom) are able to
      reproduce the results. The fit function is represented by the dotted line.
      {\bf Inset:} Blowup of the $z$-range between $-4.0$ and $-2.0\sigma$ in
      a logarithmical plot. The counterion density distribution decays
      exponentially proportional to the Gouy-Chapman length of
      $\mu=0.955\sigma$.
    \label{fig:inter_ion}
  }
\end{figure}
\begin{figure}
    \includegraphics[width=8.5cm]{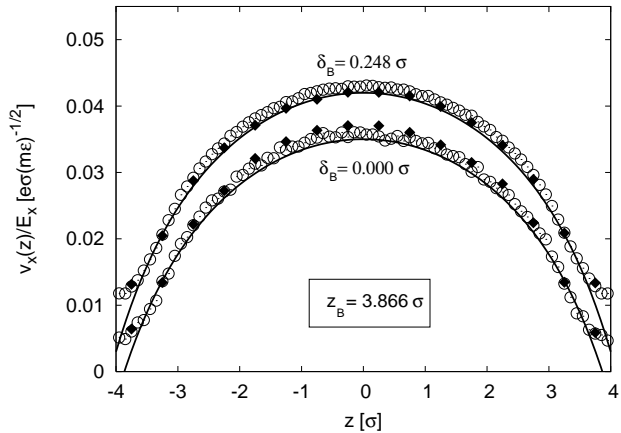}
    \caption{Counterion induced electroosmotic flow for the intermediate
      regime ($\Xi=4.19$)
      with $\rho=0.0104\sigma^{-3}$, $E_x=10.0k_B T/e\sigma$ and surface ion density
      $\sigma_s=0.042\sigma^{-2}$ with
      $\gamma_L=6.1\sigma^{-1}(m\epsilon)^{1/2}$
      ($\delta_B=(0.000\pm0.197)\sigma$) and 
      $\gamma_L=6.1\sigma^{-1}(m\epsilon)^{1/2}$ ($\delta_B=(0.248\pm0.231)\sigma$)for
      homogeneously (filled diamonds) and inhomogeneously charged walls (circles). The straight
      lines show the calculated flow profile for $|z_B|=3.866\sigma$ and for slip lengths
      as indicated.
      \label{fig:inter_EOF}
    }
\end{figure}
The simulation results for the counterion density distribution 
are presented in Fig.~\ref{fig:inter_ion}. 
The distribution can neither be fully described by the strong coupling theory 
(Eqn.~(\ref{eq:SC_density})), shown as straight line, nor by the
Poisson-Boltzmann-Theory (Eqn.~(\ref{eq:rho})), shown as dashed line.
The inset of Fig.~\ref{fig:inter_ion} shows that the counterion
distribution close to the walls decays exponentially with the Gouy-Chapman
length $\mu$, indicating the presence of a layer of highly adsorbed counterions. 
This corresponds to the behavior predicted by the strong-coupling-Theory 
(Eqn.~(\ref{eq:SC_density})).
In the bulk, however, the data are better described by the Poisson-Boltzmann 
theory in agreement with Ref.~\onlinecite{Boroudjerdi}. Thus, the intermediate regime 
bears characteristics of both the strong-coupling and the Poisson-Boltzmann
regime. For the former regime, a recently developed method \cite{Hatlo} might turn
out to be useful. 
These results are consistent with Refs.~\onlinecite{Boroudjerdi, Burak}, where it was shown 
that in case of a single charged plate, neither the Poisson-Boltzmann 
nor the strong-coupling-theory are applicable at distances from the plate
$d^{\prime}$ in the range $\sqrt{\Xi}<d^{\prime}/\mu<\Xi$. Inserting the parameters of $\Xi$ and $\mu$, 
this corresponds to the $z$-range $|z| > 2 \sigma$.

Although we have no analytical theory, we can still fit the counterion density 
with a purely heuristic function whose functional form is inspired by the prediction
of the strong coupling theory, Eq.~(\ref{eq:SC_density}),
\begin{equation}
\label{eq:test}
  \rho^t(z)=\rho_{(1)}^t\left(e^{-(z-d/2)/\mu}+e^{(z+d/2)/\mu}\right)+f_{\mbox{corr}}(z),
\end{equation}
where $\rho_{(1)}^t$ is a fitting parameter and $f_{\mbox{corr}}(z)$ a phenomenological correction function.
A good fit can be obtained with $f_{\mbox{corr}}(z) =\rho_{(2)}^t\cos(\phi z)$ 
with $\phi$ and $\rho_{(2)}^t$ being two additional fit parameters. 
The result of the fit to this expression (with fitting parameters values $\rho_{(1)}^t=(8.31\pm 0.10)\cdot 10^{-6}\sigma^{-3}$,
$\rho_{(2)}^t=(3.18\pm 0.05)\cdot 10^{-3}\sigma^{-3}$,
and $\phi=(0.416\pm 0.022)\sigma^{-1}$, respectively) 
is presented as a dotted line in Fig.~\ref{fig:inter_ion}. Combining Eq.~(\ref{eq:test}) with the Stokes 
equation, Eq.~(\ref{eq:Stokes}), and integrating it using Eq.(\ref{eq:slip}) as boundary condition, allows
to reproduce the EOF profiles very nicely (Fig.~\ref{fig:inter_EOF}).
Our result demonstrates that the Stokes equation is still applicable in the
intermediate coupling regime. Furthermore, Fig.~\ref{fig:inter_EOF} shows that
we still do not observe a noticeable effect of the surface charge inhomogeneity, 
compared to simulations with homogeneously charged walls.
Thus, electrofriction plays a negligible role in the intermediate as well as in
the weak-coupling regime
and seems to be a prominent effect only in the strong-coupling regime.
 
\section{Summary}
\label{sec:Summary}
We have carried out two types of mesoscopic simulations, namely, Dissipative Particle 
Dynamics and coupled Lattice-Boltzmann/Molecular Dynamics, of the counterion-induced EOF 
in slit channels. We have proposed a mapping scheme that allows to match quantitatively 
the parameters of the two models. The cost of calculation time for electrostatic 
problems in both methods turned out to be nearly identical. 

We have considered different electrostatic coupling regimes.
In the weak coupling regime, the numerical results for the ion density and the 
electric field are in good agreement with the predictions of the Poisson-Boltzmann theory. 
We have derived analytical expressions for the corresponding EOF profile in presence of 
no-slip as well as partial-slip boundary conditions, which also agree well with the 
simulations. 
In addition, we have considered the intermediate coupling regime where no analytical
theory is available. A heuristic function was used to fit the counterion distribution 
in this transient regime. The Stokes equation can still be used to calculate the 
corresponding EOF profiles, with very good results as compared to the simulations.

In all systems under consideration, electrofriction effects were found to be negligible.
In contrast, the presence of partial slip changes the magnitude of the flow
profiles drastically even if the slip lengths are very small \cite{Bocquet}. 
This may facilitate the generation of flow profiles in microfluidic applications. 

\section{Acknowledgements}
We thank B.~D{\"u}nweg for enlightening discussions. This work was funded by the 
Volkswagen Stiftung (Grant No.~I/80431). The simulations were carried out at the 
John-von Neumann supercomputing centre in J\"ulich (NIC), the high performance computing
center Stuttgart (HLRS) and the high performance computing cluster PC2 of
the university of Paderborn.

\end{document}